\documentclass[12pt]{article}
\usepackage{graphicx}
\usepackage{epsfig}
\usepackage{amsbsy}

\begin{document}
\baselineskip 10mm\centerline{\large \bf Thermal Stability of C$_{4+4n}$H$_{8}$ Polycubanes }

\vskip 6mm

\centerline{M. M. Maslov, A. I. Podlivaev,  and L. A. Openov$^{*}$}

\vskip 4mm

\centerline{\it National Research Nuclear University {}``MEPhI{}'', 115409 Moscow, Russia}

\vskip 2mm

$^{*}$ E-mail: LAOpenov@mephi.ru

\vskip 8mm

\centerline{\bf ABSTRACT}

The temperature dependences of the lifetime of polycubanes C$_{4+4n}$H$_{8}$ with $n=$ 2 - 5 up to their
decomposition have been directly calculated using the molecular dynamics method. It has been shown
that the activation energy of decomposition of these metastable clusters, in which the C-C bonds
form an angle of 90$^0$ that is not characteristic of carbon systems, rapidly decreases with an
increase in $n$ due to the lowering of the energy barrier that prevents the decomposition of the
clusters. This has cast some doubt on the recently made suggestion that there exist nanotubes
($n>>1$) with a square cross section. Nonetheless, the stability of bicubane ($n=2$) and tricubane
($n=3$) has proved to be sufficient for their existence at the liquid-nitrogen
temperature.

\vskip 5mm

\newpage

\centerline{\bf 1. INTRODUCTION}

In the best known carbon compounds (diamond, graphene, carbyne), the angles between covalent bonds
are equal to 109.5$^0$, 120$^0$, and 180$^0$, which correspond to the $sp^3$, $sp^2$, and $sp$
hybridizations of atomic orbitals, respectively (in fullerenes and carbon nanotubes, these angles
differ only insignificantly from 120$^0$). Consequently, increased interest has been expressed by
researchers both in the C$_8$H$_8$ cubane (Fig. 1), which was synthesized for the first time in
1964 [1] and where the angles between all adjacent C-C-C bonds are equal to 90$^0$, and in the
supercubane (Fig. 2) consisting of C$_8$ {}``cubes{}'' covalently bound to each other in the
directions of their main diagonals (note that, although there are publications on the synthesis
of the supercubane, the fact of its existence, as far as we know, has still not been proved [2, 3]).
Since so strong {}``bending{}'' of the C-C-C bonds is energetically unfavorable, the cubane and the
supercubane have a substantially strained structure. The cubane structure is stabilized by hydrogen
atoms (Fig. 1), whereas the stability of the supercubane is provided by the interaction of
neighboring elements C$_8$ (Fig. 2): in both cases, there is a saturation of {}``dangling{}'' bonds
of the carbon atoms at the vertices of the cube (an isolated cubic cluster C$_8$ is unstable [4]).

The question now arises as to the possibility of the existence of pure carbon structures that do not
contain hydrogen atoms (except, perhaps, for the passivation along the edges) and in which all C-C
bonds are either perpendicular or parallel to each other (in the supercubane, this is true only for
the C-C bonds within each individual cubic element). In [5], the polycubanes C$_{4+4n}$H$_{8}$ with
$n=$ 2 - 6 (Fig. 3) were theoretically investigated using the density functional theory method
[For polycubanes, the authors of [5] used the structural formula C$_{8+8n}$H$_8$ ($n\geq 1$). We
believe that the formula C$_{4+4n}$H$_{8}$ ($n\geq 2$) is more convenient, because, in this case,
$n$ is nothing less than the number of C$_8$ cubes in the carbon skeleton, which have (at $n>>1$)
common faces, so that $n=1$ corresponds to cubane, and $n=$ 2, 3, 4 …, to bicubane, tricubane,
tetracubane, etc.]. It was shown that all their vibration frequencies are real; i.e.,
these hydrocarbon clusters represent metastable (corresponding to local minima of the potential
energy) atomic configurations. Later [6], it was noted that, since an increase in $n$ leads to a
decrease in the relative number of hydrogen atoms in the polycubane, at $n>>1$ we are actually
dealing with a thinnest carbon nanotube that has a square cross section (hence, here, the term
{}``nanobar{}'' is more appropriate). According to [6], this nanotube can be produced under the
deformation of a conventional (2, 2) nanotube. 

It should be noted, however, that, although the real values of the vibrational frequencies
of the system provide a necessary condition for its metastability, the quantitative characteristic
of the degree of stability of the metastable state is the activation energy of its decay $E_a$,
which is involved in the Arrhenius formula for the temperature dependence of the lifetime $\tau$
of the system in this state; that is,
\begin{equation}
\tau^{-1}(T)=A\exp\left(-\frac{E_a}{k_{B}T}\right)~,
\label{1}
\end{equation}
where $A$ is the frequency factor with dimensionality s$^{-1}$, $T$ is the temperature, and $k_B$ is
the Boltzmann constant. As a rule (even though not always), the value of $E_a$ is close to the
height $U$ of the minimum energy barrier separating this metastable state from the configurations
with a lower energy. Previously [7], using the molecular dynamics method
({}``computer experiment{}''), we directly calculated the dependence of the lifetime $\tau$ on the
temperature $T$ for the C$_8$H$_8$ cubane and demonstrated that the activation energy of its
decomposition is $E_a=1.9\pm 0.1$ eV. Such a large value of $E_a$ indicates a
high thermal stability of the cubane and explains why the cubane molecules not only retain their
structure at temperatures significantly higher than room temperature [8], but also can form a
molecular crystal, the so-called solid cubane $s$-C$_8$H$_8$ with a melting temperature of
approximately 400 K [9] (during melting of this compound, only the weak van der Waals bonds between
the C$_8$H$_8$ clusters are broken, but the clusters themselves retaine their structure).

The main objective of this work is to perform the numerical simulation of the dynamics of
polycubanes C$_{4+4n}$H$_8$ with $n\ge 2$ over a wide temperature range, as well as to determine
the dependences of the activation energies of their decomposition $E_a$ and frequency factors $A$
on the number $n$. We have restricted ourselves to the first four (following the cubane)
representatives of the family: bicubane, tricubane, tetracubane, and pentacubane ($n=$ 2, 3, 4, and
5, respectively). This is enough to make sure that the thermal stability of the polycubanes
drastically decreases with increasing $n$. We have also investigated the decomposition of all the
aforementioned hydrocarbon clusters, determined the heights $U$ of the energy barriers preventing
their decomposition, and demonstrated that, within the limits of error, they coincide with their
corresponding values of $E_a$. This paper is organized as follows. Section 2 contains the
description of the computational methods. Section 3 presents the results (including the data
obtained from the molecular dynamics calculation and static simulation) and their discussion.
Section 4 provides the brief conclusion and inferences.

\vskip 10mm

\centerline{\bf 2. COMPUTATIONAL METHODS}

The energies of arbitrary atomic configurations were calculated within the framework of the
nonorthogonal tight-binding model [10], which represents a reasonable compromise between more
rigorous ab initio approaches and extremely simplified classical potentials of the interatomic
interactions. This model adequately describes the structural and energy characteristics of various
hydrocarbon systems, including both small clusters (cubane [7], methylcubane [11], etc.), and
macroscopic systems (graphane [12], graphone [13], hydrogen interstitials in diamonds [10], etc.).
Yielding to the accuracy of the ab initio methods, this model requires a considerably less
expenditure of computer resources and, therefore, makes it possible to investigate the evolution
of the system consisting of 10 - 100 atoms for a time (from 1 ns to 1 $\mu$s) sufficient to recruit
the necessary statistics [14]. Previously, we, in particular, successfully used it to simulate
the spontaneous regeneration of the graphene/graphane interface during thermal disordering [15],
to determine the dependence of the dielectric gap of graphane nanoribbons from their width [16],
etc.

For the simulation of the dynamics of polycubanes, at the initial instant of time, random velocities
and displacements were imparted to each of the atoms in such a manner that the momentum and the
angular momentum of the cluster as a whole were equal to zero. Then, the forces acting on the atoms
were calculated using the tight-binding model and the Hellmann-Feynman theorem [10]. The classical
Newton equations of motion were numerically integrated using the velocity Verlet method with the
time step $t_{0}=2.72\times 10^{-16}$ s [14]. In the course of the simulation, the total energy of
the cluster (the sum of the potential and kinetic energies) remained unchanged, which corresponds
to a microcanonical ensemble (the system is not in a thermal equilibrium with the environment
[7, 11-15]). In this case, the {}``dynamic temperature{}'' $T$ is a measure of the energy of
relative motion of the atoms and is calculated from the formula [17, 18]
$\langle E_{\textrm{kin}} \rangle=\frac{1}{2}k_{B}T(3n-6)$, where
$\langle E_{\textrm{kin}} \rangle$ is the time-averaged kinetic energy of the cluster and $n$ is the
number of atoms in the cluster (here, it is taken into account that the cluster as a whole neither
moves nor rotates; therefore, the number of degrees of freedom is reduced by six). 

In order to determine the heights $U$ of the energy barriers preventing the decomposition of the
polycubanes, we investigated the hypersurface of the potential energy $E_{pot}$ for each value of $n$ as a function of the atomic coordinates in the vicinity of the initial metastable
configuration and used the method of searching for the saddle point nearest to the local minimum
of the potential energy $E_{pot}$ in normal coordinates (for more details, see [19, 20]). It should
be noted that the task of finding the saddle point in this case can be substantially simplified if
we use the molecular dynamics data, because they provide information both about the isomer, into
which one or another polycubane transforms upon the decomposition, and about the atomic
configurations that arise in the course of this transformation.

\vskip 10mm

\centerline{\bf 3. RESULTS AND DISCUSSION}

The binding energies of atoms in the polycubanes C$_{4+4n}$H$_8$, which were calculated according
to the formula [10] 
\begin{equation}
E_b=\frac{(4+4n)E(\mathrm{C})+8E(\mathrm{H})-E(\mathrm{C}_{4+4n}\mathrm{H}_8)}{12+4n}
\label{2}
\end{equation}
are equal to 4.51, 4.59, 4.65, and 4.69 eV/atom for $n=$ 2, 3, 4, and 5, respectively. For
comparison, the binding energy in the C$_8$H$_8$ cubane is $E_b=4.42$ eV/atom [7]. Thus, the
binding energy $E_b$ increases with increasing $n$, which indicates an increase in the
thermodynamic stability. As will be shown below, this is not accompanied by an increase in the
kinetic stability. 

For each $n=$ 2 - 5, we investigated the evolution of the C$_{4+4n}$H$_8$ cluster for 30 - 60
different sets of initial velocities and displacements of atoms, which correspond to the temperature
ranges $T=$ 500 - 1700 K ($n=2$), 300 - 1400 K ($n=3$), 80 - 500 K ($n=4$), and 50 - 200 K ($n=5$).
The temperature range every time was chosen in such a way that the lifetime $\tau$ of the
corresponding polycubane up to its decomposition would change in the interval from $\sim 1$ ps to
$\sim 0.1$ $\mu$s, which corresponds to molecular dynamics steps in the range from
$\sim 3\times 10^3$ to $\sim 3\times 10^8$, respectively. The choice
$\tau \sim 1$ ps as the lower boundary of this interval is motivated by the requirement
$\tau >> \tau_0$, where $\tau_0\sim 30$ fs is the period of the fastest vibrations of the
C$_{4+4n}$H$_8$ clusters. The upper boundary of the aforementioned interval $\tau\sim 0.1$ $\mu$s
was chosen rather arbitrarily and is dictated by the limited speed of the computers (for the
C$_8$H$_8$ cubane, it can be increased to $\sim 1$ $\mu$s [7]). As will be shown below, the
significant difference in the temperature intervals (and in the mean values of temperatures in
these intervals) at different values of $n$ is associated with the different thermal stabilities of
the C$_{4+4n}$H$_8$ polycubanes studied in our work.

Figure 4 presents the results of the direct numerical calculation of the lifetime $\tau$ at
different temperatures $T$ for bicubane ($n=2$), tricubane ($n=3$), and tetracubane ($n=4$). The
data obtained for pentacubane ($n=5$) do not present in this figure for the purpose of facilitating
its perception. It can be seen that the dependences of ln$\tau$ on 1/$T$ for all values of $n$ are
well approximated by straight lines in accordance with the Arrhenius equation (1) without a
correction for the finite sizes of the thermal reservoir [21-23]. The slopes of these lines
determine the corresponding activation energies $E_a$, whereas the points of their intersection
with the ordinate axis determine the frequency factors $A$ involved in formula (1). A statistical
analysis of the data obtained from the {}``computer experiment{}''  gives the following values of
the aforementioned parameters: $E_a=0.76\pm 0.1$, $0.36\pm 0.05$, $0.10\pm 0.02$, and $0.07\pm 0.02$
eV and $A=10^{14.79\pm 0.22}$, $10^{14.16\pm 0.12}$, $10^{13.05\pm 0.14}$, $10^{13.72\pm 0.72}$
s$^{-1}$ for $n=2$, 3, 4, and 5, respectively. Taking into account that $E_a=1.9\pm 0.1$ eV for the
C$_8$H$_8$ cubane [7], we can draw the conclusion that the activation energy $E_a$ of decomposition
of the polycubanes in the C$_{4+4n}$H$_8$ family monotonically decreases with increasing $n$
(Fig. 5). The observed decrease in the stability of the polycubanes is apparently associated with
the fact that, beginning with  $n=2$ in the C$_{4+4n}$H$_8$ clusters, there arise purely carbon
C-C-C bonds, which are bent at a right angle and do not contain (unlike the case of the C$_8$H$_8$
cubane) hydrogen atoms. It can be seen from Fig. 4 that, for each value of $n$, the lifetime $\tau$
exponentially decreases with an increase in the temperature $T$ and that, at a fixed temperature
$T < 1000$ K, it decreases (also exponentially) with an increase in the number $n$ of C$_8$ cubes
in the carbon skeleton of the polycubane. 

The frequency factor A with increasing number $n$ also exhibits a tendency toward a decrease
(according to the data reported in [7] for the C$_8$H$_8$ cubane, the frequency factor is equal to
$10^{16.03\pm 0.36}$ s$^{-1}$). Consequently, at a sufficiently high temperature $T > 1000$ K,
the lifetime of the polycubanes increases with an increase in their length, because, owing to the
small ratio $E_a/k_B T$, the preexponential factor in the Arrhenius equation (1) does not play a
decisive role. Note, however, that interest expressed in the range of these temperatures is rather
limited, because the absolute value of $\tau$ in this case remains very small ($\sim 1$ ps, Fig. 4). 

Let us analyze the structure and energy of the isomers formed upon the decomposition of the
polycubanes. The molecular dynamics data indicate that the decomposition of the C$_{12}$H$_8$
bicubane almost always occurs as a result of the synchronous breaking of two C-C bonds, which are
directed perpendicular to the bicubane axis (and parallel to each other) and located at the center
of the cluster. This results in the formation of an isomer shown in Fig. 6. Two opposite side
faces of this isomer are hexagons formed by the C-C bonds, as is the case in nanotubes. The binding
energy in the isomer increases to $E_b=4.68$ eV/atom; i.e., such isomer is energetically more
favorable as compared to the bicubane (see formula (2)). In the further evolution, this isomer
transforms into different quasi-two-dimensional clusters and no return to the initial configuration
of the bicubane occurs. 

The decomposition of the C$_{16}$H$_8$ tricubane begins with the sequential formation of the isomers
depicted in Fig. 7. First, one pair of C-C bonds perpendicular to the tricubane axis (and parallel
to each other) undergoes breaking, followed by the breaking of another pair of bonds that is
perpendicular to the first pair and shifted relative to it along the tricubane axis by the length of
the C-C bond. This leads to the formation of a cluster containing four hexagons composed of C-C
bonds. The binding energy $E_b=5.06$ eV/atom of this cluster is higher than that of the tricubane.
Then, this isomer transforms into different low-dimensional hydrocarbon complexes. We have never
observed the recovery of the initial tricubane.

The decomposition of the polycubanes C$_{4+4n}$H$_8$ with $n > 3$ occurs in a similar manner. Every time,
after the synchronous breaking of pairs of transverse C-C bonds, the rectangles consisting of C-C
bonds sequentially transform into hexagons so that, at $n >> 1$, we eventually observe the formation
of the (2,2) nanotube passivated with hydrogen atoms along the edges (see also [6]).

Since the breaking of the first pair of C-C bonds leads to an irreversible decomposition of each
polycubane, the height $U$ of the energy barrier preventing this decomposition represents just the
height of the barrier to the decomposition of this polycubane. Figure 8 presents the results of the
calculation of the potential energy $E_{pot}$ of the C$_{12}$H$_8$ bicubane along the reaction
coordinate connecting the initial metastable configuration with the configuration of the isomer
shown in Fig. 6. The energy barrier height thus determined, i.e., $U=0.69$ eV, coincides within the
statistical error with the activation energy of thermally activated decomposition of the bicubane
$E_a=0.76\pm 0.1$ eV. Similarly, we have determined the height $U=0.39$ eV of the energy barrier to
the decomposition of the C$_{16}$H$_8$ tricubane (Fig. 9). This value is also consistent with the
corresponding activation energy $E_a=0.36\pm 0.05$ eV. It should be noted that, for the first
isomer formed from the tricubane, there is a very shallow local minimum of the potential energy
$E_{pot}$ with a depth of $\sim 0.01$ eV (Fig. 9). Consequently, in the molecular dynamics
simulation of the evolution of the tricubane, we have observed configurations that are very close
to this isomer for a very short time ($\sim 1$ ps), followed by the formation of the second isomer
C$_{16}$H$_8$ (Fig. 7b). For tetracubane and pentacubane, we found the energy barrier heights
$U=0.11$ and 0.06 eV, respectively, which is also consistent with the activation energies
$E_a=0.10\pm 0.02$ and $0.07\pm 0.02$ eV, respectively.

\vskip 10mm

\centerline{\bf 4. CONCLUSIONS}

Thus, the results presented in this paper on the numerical simulation of the dynamics of
polycubanes C$_{4+4n}$H$_8$ with different values of $n$ indicate a rapid decrease in their
thermal stability with an increase in the number $n$ of cubic fragments C$_8$ in the carbon
skeleton. For each member of the family of polycubanes with  $n=$ 2 - 5, we have determined the
activation energy of its decomposition $E_a$ and the frequency factor $A$, which have allowed one
to find the lifetimes $\tau$ at an arbitrary temperature. In particular, at $T=300$ K, we have
$\tau \sim 10$ ms, 10 ns, 10 ps, and 0.1 ps for bicubane, tricubane, tetracubane, and pentacubane,
respectively; i.e., an increase in the length of the polycubane leads to a rapid decrease in $\tau$.
Therefore, it is hardly possible to conclude that, under normal conditions, there exist
{}``nanotubes{}'' with a square cross section (even if their edges are passivated with hydrogen).

Meanwhile, for the lifetime of bicubane and tricubane at a temperature of 77 K, we have the
estimates $\tau \sim 10^{35}$ and 10$^9$ s, respectively. This suggests that the possibility
exists of monitoring these unique hydrocarbon clusters at cryogenic temperatures (here, we do not
touch the problem of their synthesis, which is a separate and very difficult task). Of particular
interest (by analogy with solid cubane $s$-C$_8$H$_8$) is the preparation of anisotropic solid
bicubane $s$-C$_{12}$H$_8$ and solid tricubane $s$-C$_{16}$H$_8$, in which polycubanes oriented in
one direction are bound to each other by means of van der Waals bonds. It is also  interesting to
consider the existence of a superbicubane and a supertricubane, in which structural elements are
rectangular parallelepipeds C$_{12}$ and C$_{16}$, respectively.

\newpage
\vskip 2mm
\includegraphics[width=\hsize,height=13cm]{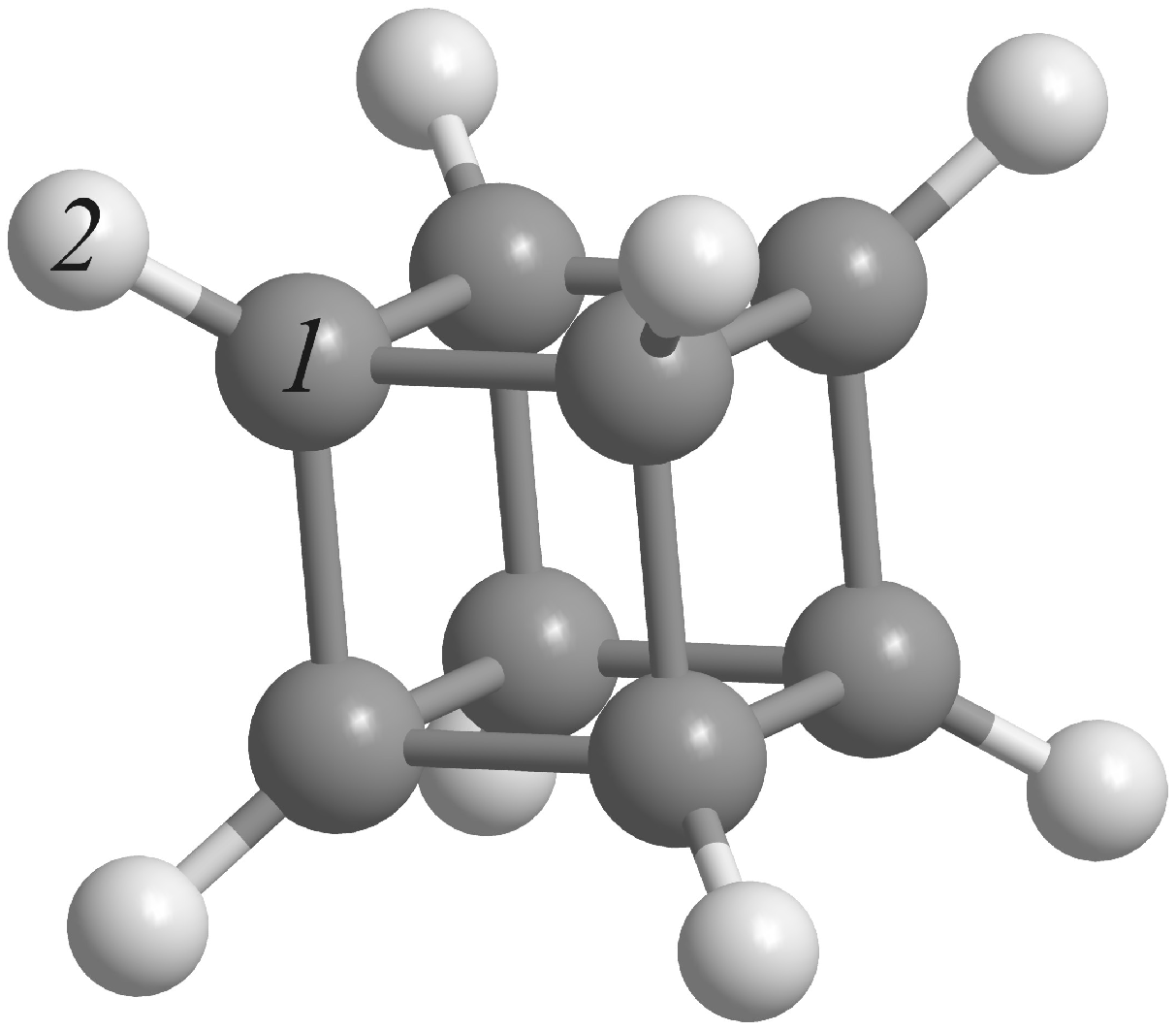}
\vskip 20mm
Fig. 1. C$_8$H$_8$ cubane: (1) carbon atoms and (2) hydrogen atoms. 

\newpage
\vskip 2mm
\includegraphics[width=\hsize,height=15cm]{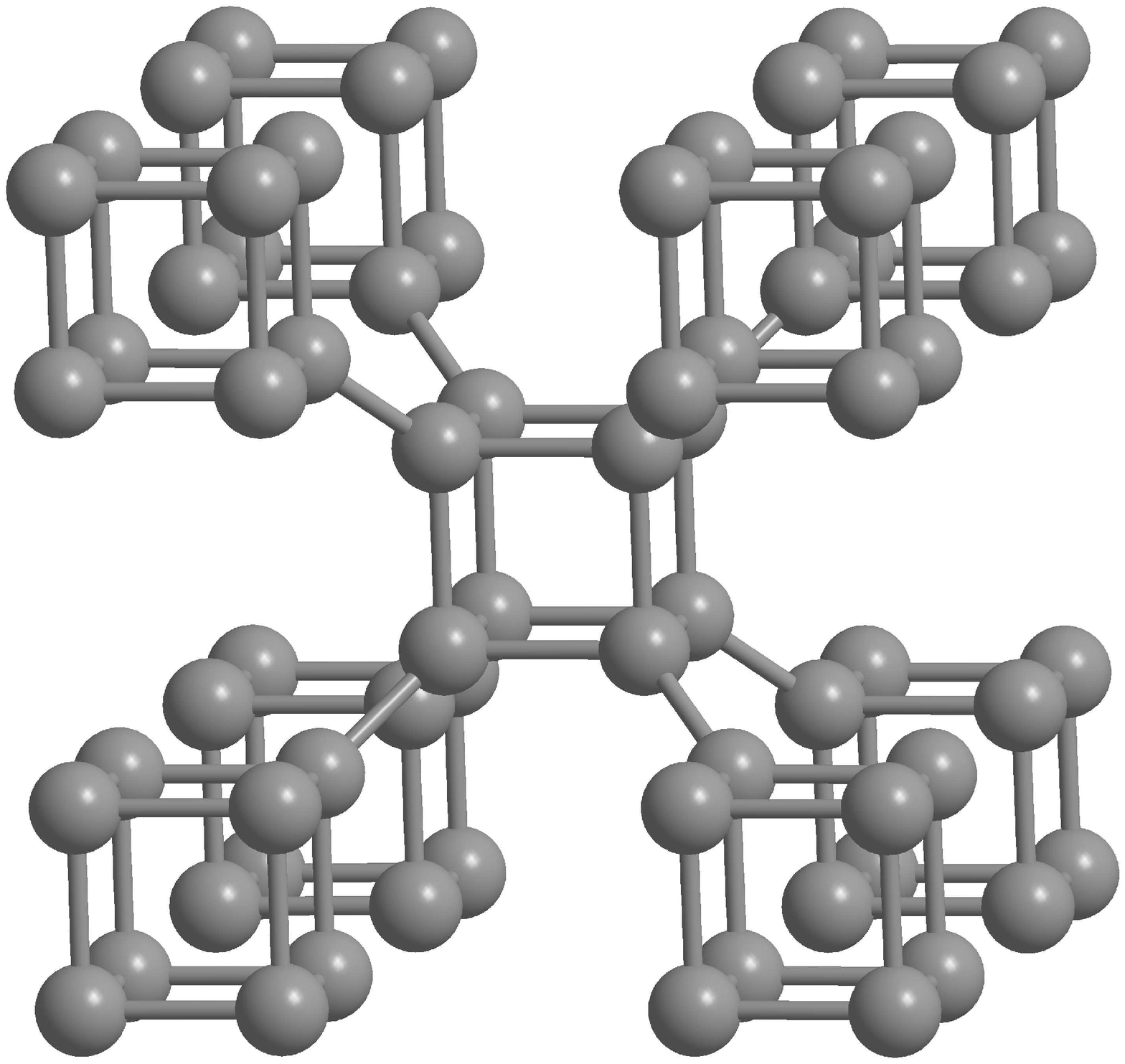}
\vskip 20mm
Fig. 2. A fragment of the supercubane.

\newpage
\vskip 2mm
\includegraphics[width=\hsize,height=12cm]{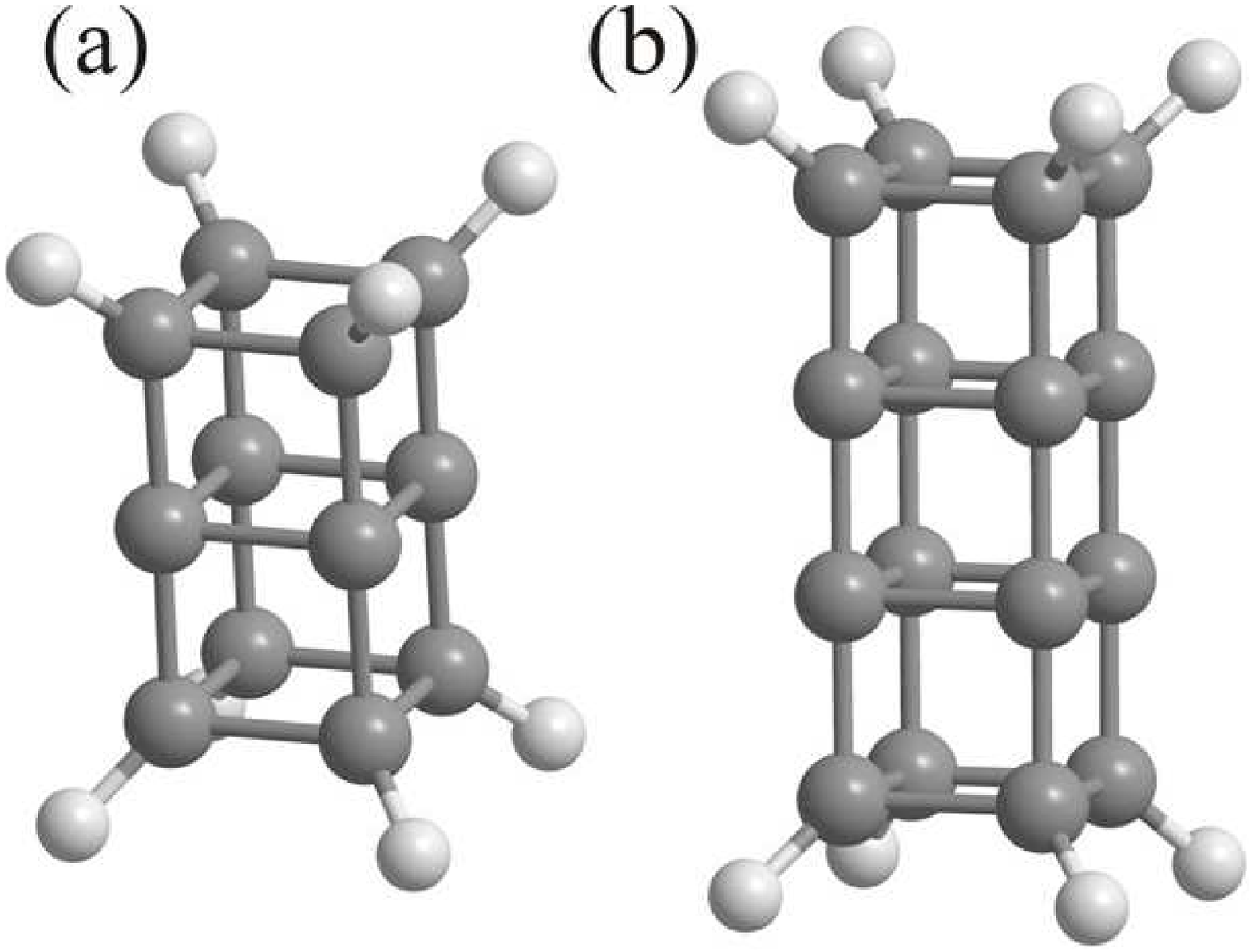}
\vskip 20mm
Fig. 3. (a) Bicubane C$_{12}$H$_8$ and (b) tricubane C$_{16}$H$_8$. The
notation of the atoms is the same as that used in Fig. 1. 

\newpage
\vskip 2mm
\includegraphics[width=\hsize,height=14cm]{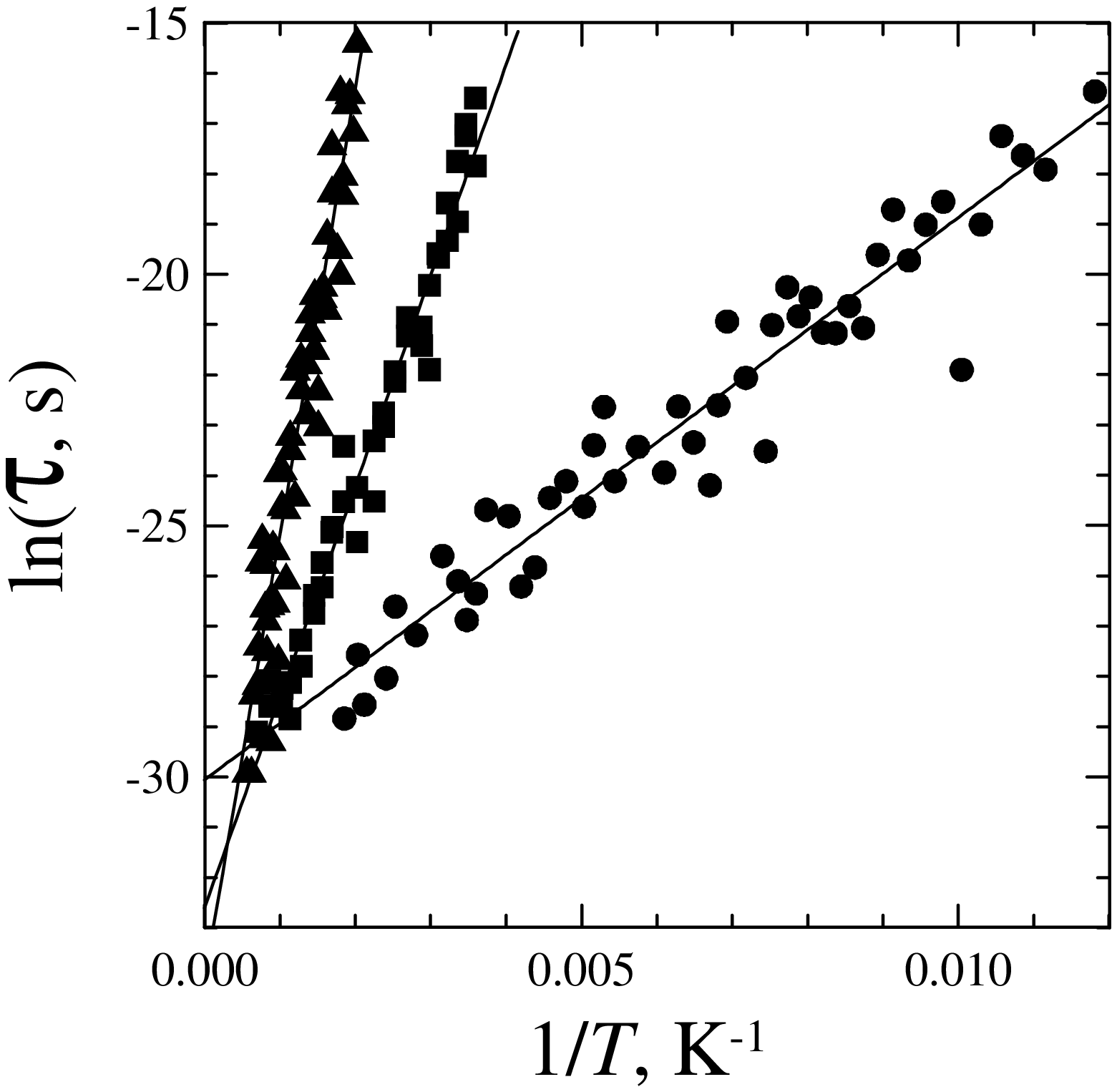}
\vskip 20mm
Fig. 4. Dependences of the logarithm of the lifetime $\tau$ for the (1) bicubane C$_{12}$H$_8$,
(2) tricubane C$_{16}$H$_8$, and (3) tetracubane C$_{20}$H$_8$ on the reciprocal of the initial
temperature $T$. Solid lines are the corresponding linear approximations obtained by the
least-squares method.

\newpage
\vskip 2mm
\includegraphics[width=\hsize,height=15cm]{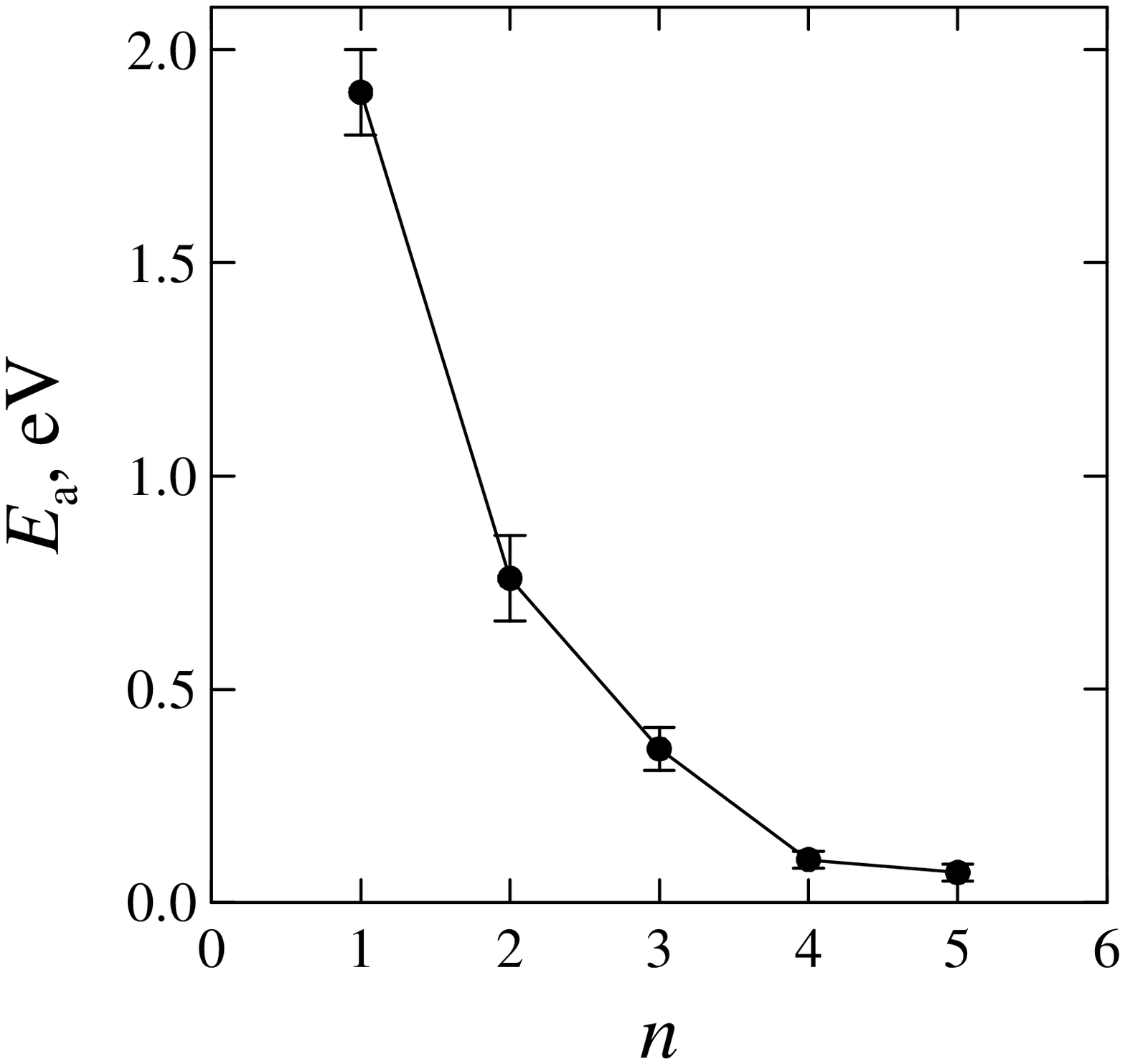}
\vskip 20mm
Fig. 5. Dependence of the activation energy of decomposition of the C$_{4+4n}$H$_8$ cluster on the
number $n$ of C$_8$ cubes in the carbon skeleton (data for $n=1$ are taken from [7]).

\newpage
\vskip 2mm
\includegraphics[width=12cm,height=15cm]{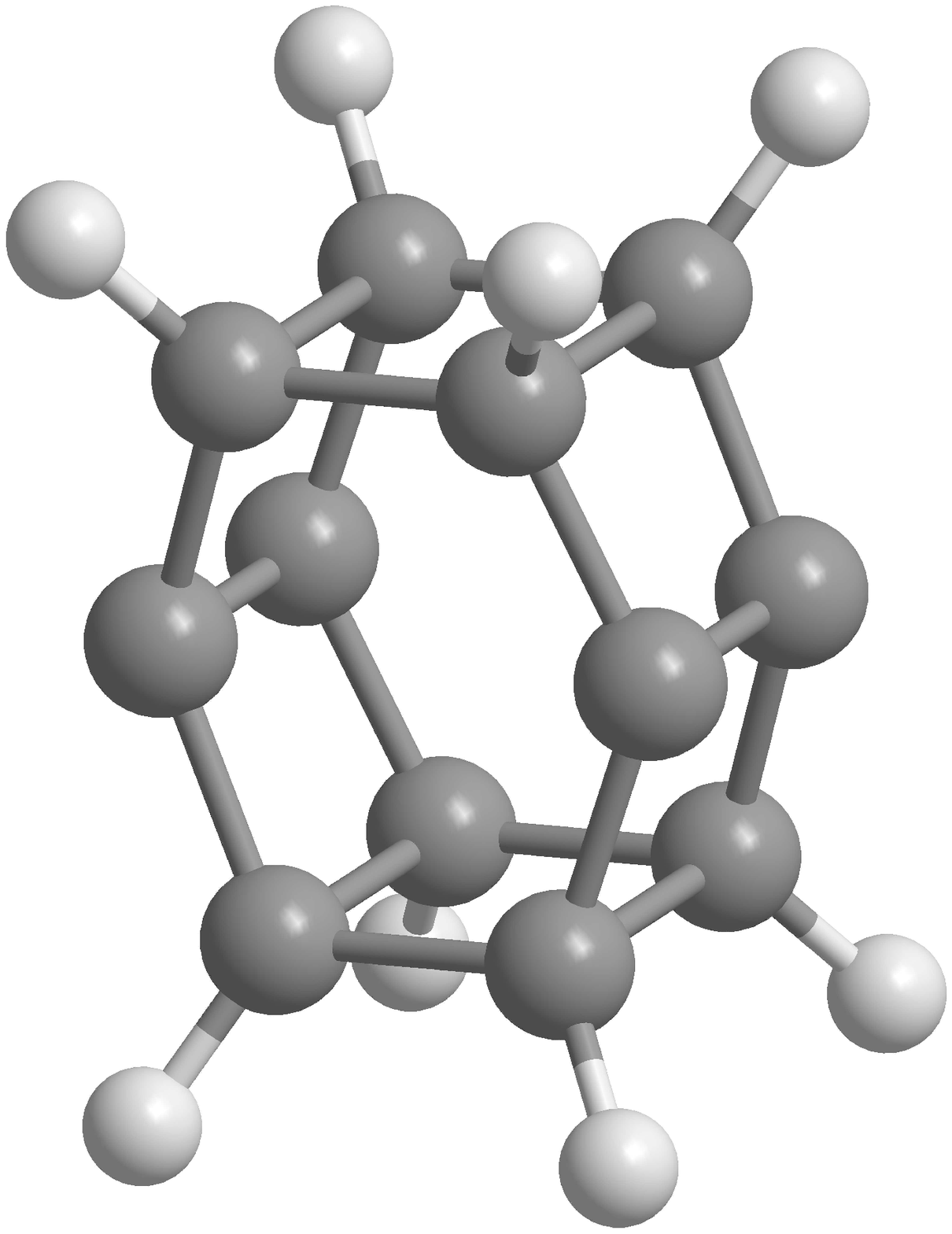}
\vskip 20mm
Fig. 6. Isomer C$_{12}$H$_8$ into which the bicubane transforms upon the decomposition. The notation
of the atoms is the same as that used in Fig. 1. 

\newpage
\vskip 2mm
\includegraphics[width=\hsize,height=14cm]{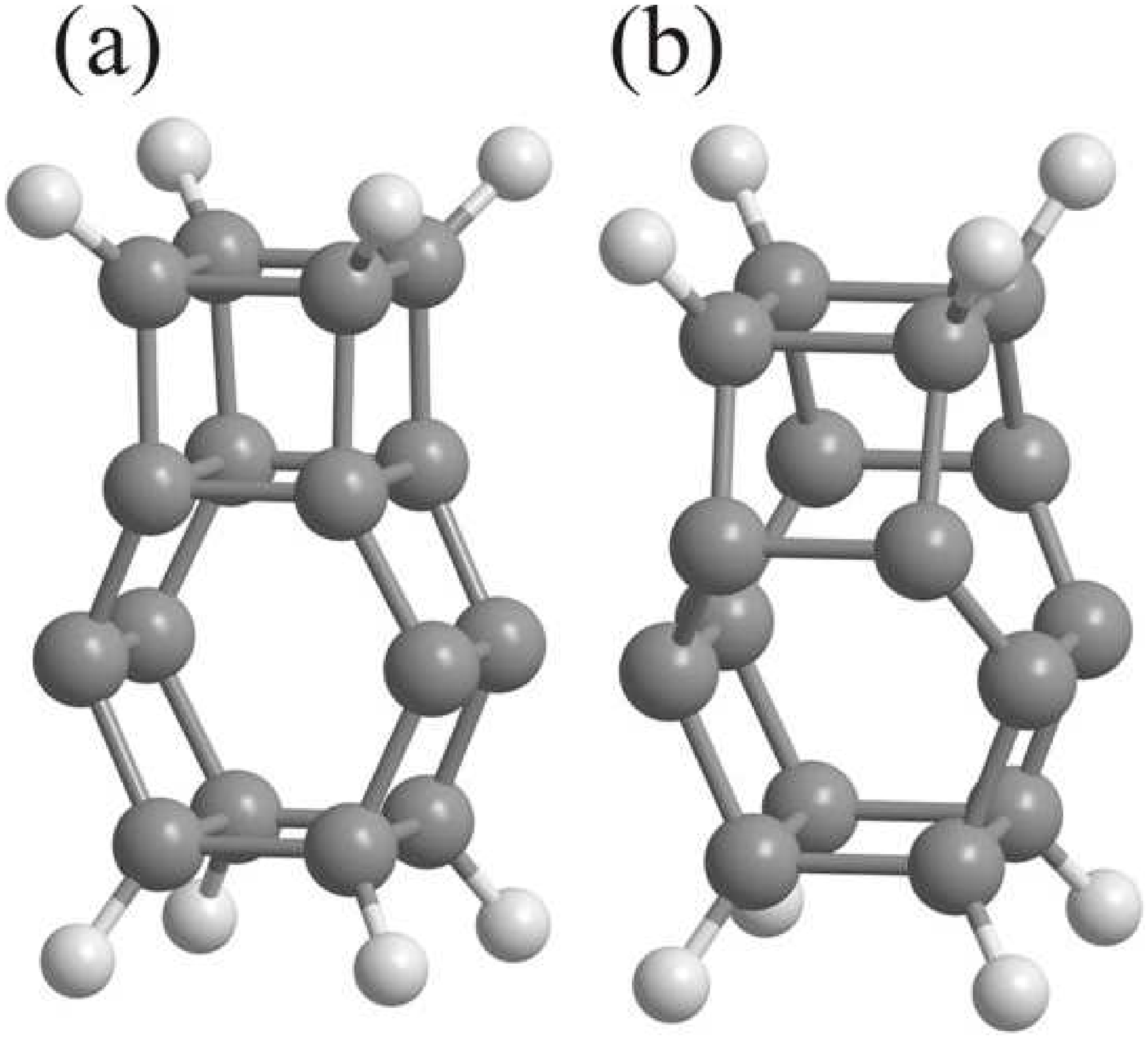}
\vskip 20mm
Fig. 7. Isomers C$_{16}$H$_8$ into which the tricubane sequentially transforms upon the
decomposition: (a) metastable isomer and (b) low-energy isomer. The notation of the atoms is the
same as that used in Fig. 1. 

\newpage
\vskip 2mm
\includegraphics[width=\hsize,height=15cm]{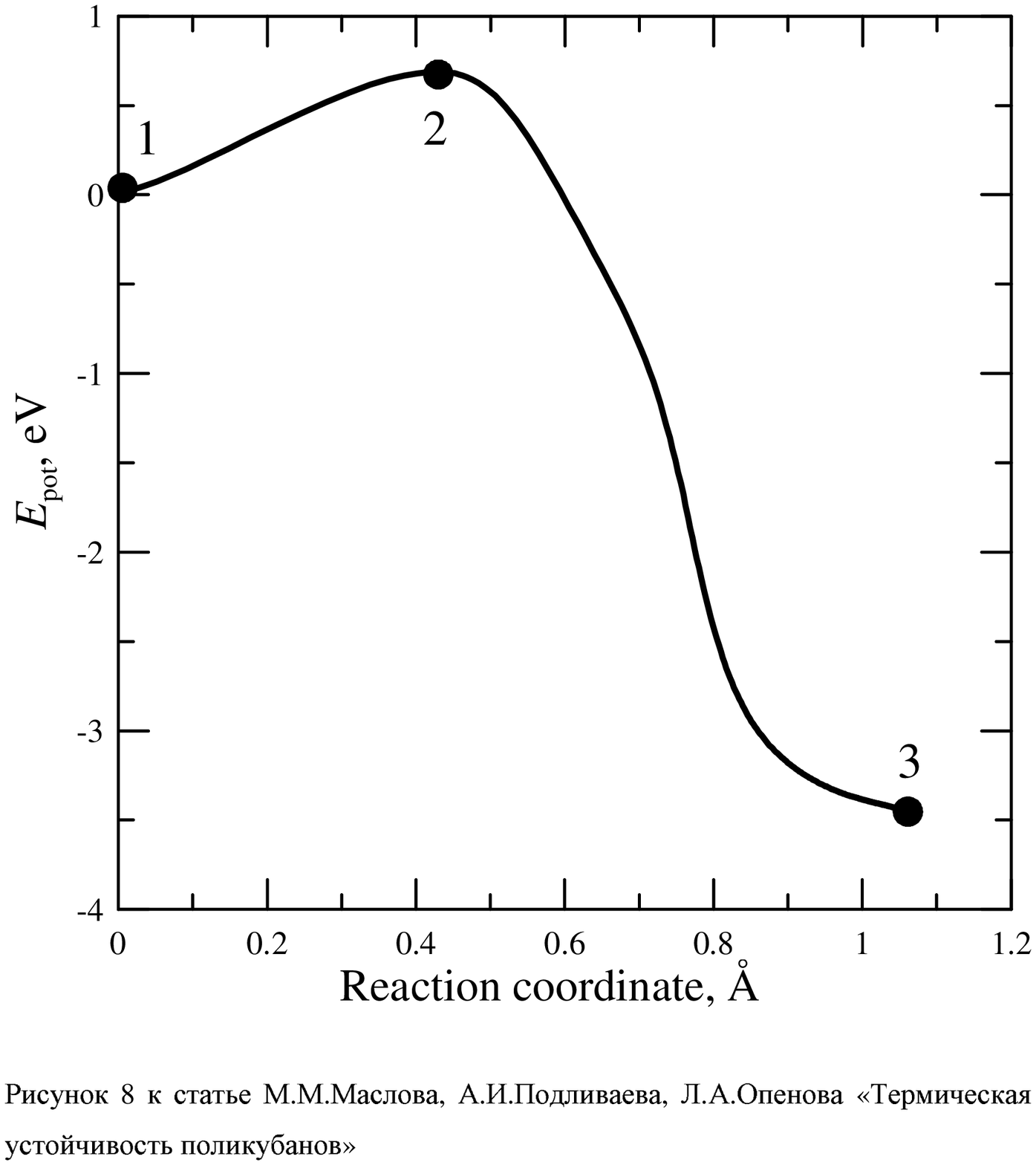}
\vskip 20mm
Fig. 8. Potential energy $E_{pot}$ of the C$_{12}$H$_8$ cluster in the vicinity of the configuration
of the bicubane: (1) bicubane (Fig. 3a), (2) saddle point that determines the barrier $U=0.69$ eV
for the decomposition of the bicubane, and (3) low-energy isomer (Fig. 6). The reaction coordinate
is chosen as the half the sum of increments in the lengths of two C-C bonds broken during the
decomposition of the bicubane.

\newpage
\vskip 2mm
\includegraphics[width=\hsize,height=15cm]{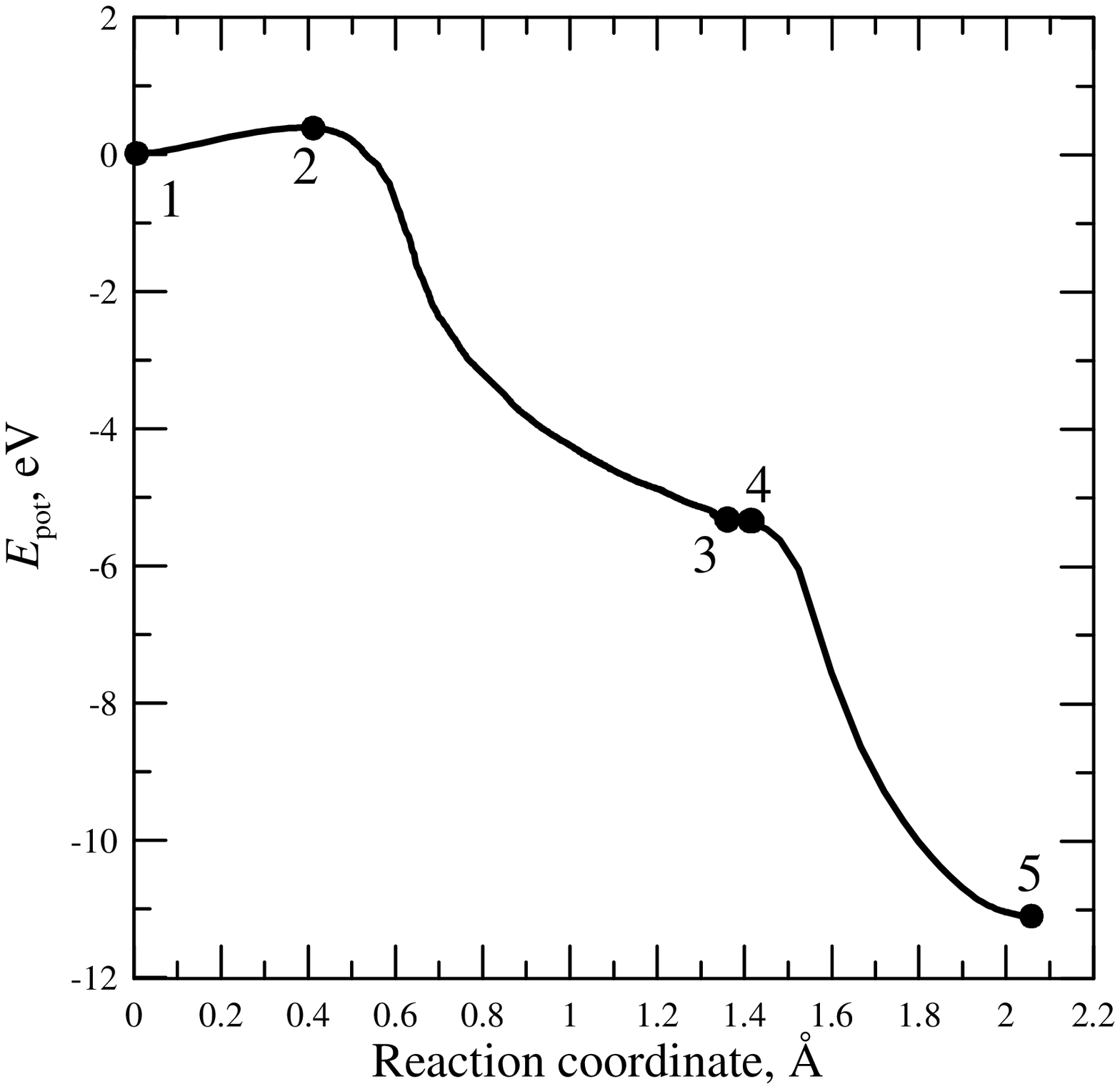}
\vskip 5mm
Fig. 9. Potential energy $E_{pot}$ of the C$_{16}$H$_8$ cluster in the vicinity of the configuration
of the tricubane: (1) tricubane (Fig. 3b), (2) saddle point that determines the barrier $U=0.39$ eV
for the decomposition of the tricubane, (3) metastable isomer (Fig. 7a), (4) saddle point that
determines the barrier $U=0.008$ eV for the decomposition of this isomer, and (5) low-energy isomer
(Fig. 7b). The reaction coordinate is chosen as the half the sum of increments in the lengths of
four C-C bonds sequentially broken during the decomposition of the tricubane (see text). 

\end{document}